\documentclass[letterpaper,twoside,twocolumn,aps,prl,floatfix,superscriptaddress,showpacs]{revtex4}
\usepackage{graphics}
\usepackage{hyperref}

\begin{document}

\title{$ J/\psi  $ Polarization in 800-GeV $ p- $Cu Interactions}

\affiliation{Abilene Christian University, Abilene, TX 79699}
\affiliation{Argonne National Laboratory, Argonne, IL 60439}
\affiliation{Fermi National Accelerator Laboratory, Batavia, IL 60510}
\affiliation{Georgia State University, Atlanta, GA 30303}
\affiliation{Illinois Institute of Technology, Chicago, IL  60616}
\affiliation{Los Alamos National Laboratory, Los Alamos, NM 87545}
\affiliation{New Mexico State University, Las Cruces, NM, 88003}
\affiliation{Oak Ridge National Laboratory, Oak Ridge, TN 37831}
\affiliation{Texas A\&M University, College Station, TX 77843}
\affiliation{Valparaiso University, Valparaiso, IN 46383}

\author{T.H.~Chang}
\altaffiliation[Present address: ]{University of Illinois, Urbana, IL 61801}
\affiliation{New Mexico State University, Las Cruces, NM, 88003}

\author{M.E.~Beddo}
\affiliation{New Mexico State University, Las Cruces, NM, 88003}

\author{C.N.~Brown}
\affiliation{Fermi National Accelerator Laboratory, Batavia, IL 60510}

\author{T.A.~Carey}
\affiliation{Los Alamos National Laboratory, Los Alamos, NM 87545}

\author{W.E.~Cooper}
\affiliation{Fermi National Accelerator Laboratory, Batavia, IL 60510}

\author{C.A.~Gagliardi}
\affiliation{Texas A\&M University, College Station, TX 77843}

\author{G.T.~Garvey}
\affiliation{Los Alamos National Laboratory, Los Alamos, NM 87545}

\author{D.F.~Geesaman}
\affiliation{Argonne National Laboratory, Argonne, IL 60439}

\author{E.A.~Hawker}
\altaffiliation[Present address: ]{University of Cincinnati, Cincinnati, OH 45221}
\affiliation{Texas A\&M University, College Station, TX 77843}
\affiliation{Los Alamos National Laboratory, Los Alamos, NM 87545}

\author{X.C.~He}
\affiliation{Georgia State University, Atlanta, GA 30303}

\author{L.D.~Isenhower}
\affiliation{Abilene Christian University, Abilene, TX 79699}

\author{D.M.~Kaplan}
\affiliation{Illinois Institute of Technology, Chicago, IL  60616}

\author{S.B.~Kaufman}
\affiliation{Argonne National Laboratory, Argonne, IL 60439}

\author{D.D.~Koetke}
\affiliation{Valparaiso University, Valparaiso, IN 46383}


\author{P.L.~McGaughey}
\affiliation{Los Alamos National Laboratory, Los Alamos, NM 87545}

\author{W.M.~Lee}
\altaffiliation[Present address: ]{Florida State University, Tallahassee, FL 32306}
\affiliation{Georgia State University, Atlanta, GA 30303}

\author{M.J.~Leitch}
\affiliation{Los Alamos National Laboratory, Los Alamos, NM 87545}

\author{J.M.~Moss}
\affiliation{Los Alamos National Laboratory, Los Alamos, NM 87545}

\author{B.A.~Mueller}
\affiliation{Argonne National Laboratory, Argonne, IL 60439}

\author{V.~Papavassiliou}
\affiliation{New Mexico State University, Las Cruces, NM, 88003}

\author{J.C.~Peng}
\altaffiliation[Present address: ]{University of Illinois, Urbana, IL 61801}
\affiliation{Los Alamos National Laboratory, Los Alamos, NM 87545}


\author{P.E.~Reimer}
\affiliation{Los Alamos National Laboratory, Los Alamos, NM 87545}
\affiliation{Argonne National Laboratory, Argonne, IL 60439}

\author{M.E.~Sadler}
\affiliation{Abilene Christian University, Abilene, TX 79699}

\author{W.E.~Sondheim}
\affiliation{Los Alamos National Laboratory, Los Alamos, NM 87545}

\author{P.W.~Stankus}
\affiliation{Oak Ridge National Laboratory, Oak Ridge, TN 37831}

\author{R.S.~Towell}
\affiliation{Abilene Christian University, Abilene, TX 79699}
\affiliation{Los Alamos National Laboratory, Los Alamos, NM 87545}

\author{R.E.~Tribble}
\affiliation{Texas A\&M University, College Station, TX 77843}

\author{M.A.~Vasiliev}
\altaffiliation[On leave from ]{Kurchatov Institute, Moscow, Russia}
\affiliation{Texas A\&M University, College Station, TX 77843}

\author{J.C.~Webb}
\altaffiliation[Present address: ]{Indiana University, Bloomington, IN 61801}
\affiliation{New Mexico State University, Las Cruces, NM, 88003}

\author{J.L.~Willis}
\affiliation{Abilene Christian University, Abilene, TX 79699}

\author{G.R.~Young}
\affiliation{Oak Ridge National Laboratory, Oak Ridge, TN 37831}

\collaboration{FNAL E866/NuSea Collaboration}
\noaffiliation

\date{\today}

\begin{abstract}
We present measurements of the polarization of the $ J/\psi  $
produced in 800-GeV proton interactions with a copper target. Polarization
of the $ J/\psi $ is sensitive to the $ c \overline{c} $ production and
hadronization processes. A longitudinal
polarization is observed at large $ x_{F} $, while at small $ x_{F} $
the state is produced essentially unpolarized or slightly transversely
polarized. No significant variation of the polarization is observed versus 
$ p_{T} $.
\end{abstract}



\maketitle

A detailed understanding of the production mechanism of charmonium is
important for ongoing research at the Relativistic Heavy Ion Collider,
where this process can play an important role in the search for
quark-gluon plasma formation~\cite{Matsui:1986dk}, as well as for the
investigation of the gluon contribution to the proton spin
structure~\cite{Robinett:1991cd,Teryaev:1997sr}. In the non-relativistic
QCD formalism (NRQCD), charmonium polarization can probe~\cite{Bodwin:1995jh} 
details of the production process~\cite{Cho:1995ih} that are
perturbatively calculable.

NRQCD is an effective field theory that approximates the full QCD
Lagrangian for large quark masses. Uncalculable matrix elements for the
production of various Fock-space components of the quarkonium wavefunction
are ranked, using simple scaling rules, according to their order in $ v $,
the relative quark-antiquark velocity in the quarkonium rest frame; for
charmonium, this has a value of $ \sim0.5 $. The leading matrix elements
can then be extracted from fits to experimental data and used in
calculating other processes.

A determination~\cite{Kniehl:1998qy} of the various matrix elements from
available high-energy data defines to a large extent the expected
production properties of charmonium, one of which is the polarization. For
a spin-1 particle, given the fraction $\xi$ of particles produced in the
$j_z=0$ (``longitudinal'') state, we can define the polarization
$\lambda =(1-3\xi )/(1+\xi ) $; $\lambda$ is positive (negative)
for transverse (longitudinal) polarization. 
While several of the intermediate $ c\overline{c} $ states
are color-octet states and must be followed by multiple gluon emission
before a physical charmonium state is produced, heavy-quark symmetry
implies that gluon radiation leaves the quark spins unchanged, providing
definite predictions for the spin state of the final 
charmonium~\cite{Cho:1995ih}.

The polarization of the $ J/\psi  $ has been measured with relatively
high statistical precision only in fixed-target experiments, in 
pion~\cite{Biino:1987qu} and proton~\cite{Alexopoulos:1997yd,Gribushin:1999ha} 
interactions with solid nuclear targets.
No significant polarization has been seen in either, with the exception
of an intriguing large longitudinal polarization at the highest $ x_{F} $
value of the pion-induced data. Measurements in collider experiments, at 
energies and transverse momenta where theoretical calculations should be 
more robust and which are free of potential complications from nuclear
effects, suffer from low statistics~\cite{Affolder:2000nn}. Finally, in a 
recent study of fixed-target
bottomonium production~\cite{Brown:2001bz}, the $ \Upsilon (1S) $  was 
found to be largely unpolarized, while the $ \Upsilon (2S,3S) $ states had 
strong transverse polarization.

In this experiment, we accumulated~\cite{Chang:1999hd} a much larger 
sample --- approximately 9 million --- of reconstructed $ J/\psi  $
than any previous study, in interactions of an 800-GeV/$ c $ proton
beam with a copper target. $ J/\psi  $ decays were measured 
using the Fermilab Meson-East dimuon spectrometer~\cite{Moreno:1991sf},
which consisted of three dipole magnets, SM0, SM12, and SM3, and three
stations of drift chambers and trigger scintillator hodoscopes. Data
were collected during a month-long dedicated run in which the copper
beam dump inside SM12 was used in place of a target; the first
dipole was switched off during this run. A copper absorber filtered
out all hadrons and electrons produced in the interactions, allowing
only muons to enter the spectrometer. Additional hadron rejection
was obtained using a muon identifier, consisting of proportional tubes
and scintillator hodoscopes behind a thick absorber at the downstream 
end of the apparatus. Events were recorded when the trigger condition 
of two oppositely-charged muons was satisfied.

Events were reconstructed offline from the recorded hits in the drift
chambers. Tracks were traced back through the magnetic fields to the
dump/target, where a vertex was formed, consistent with the beam position.
Energy losses and multiple Coulomb scattering in the absorber and
the dump were taken into account in the traceback. The momentum of
the muons was determined by their bending in SM3 and this was used
to calculate their trajectories in SM12.

The invariant mass of the muon pair was calculated from the muon momenta
and opening angle at the vertex. Since the energy loss and multiple
scattering are only known on average and not on an event-by-event
basis, the mass resolution, typically $ \approx 500 $~MeV (FWHM), 
was not sufficient to separate the $ J/\psi  $
and $ \psi ^{\prime } $ peaks in this run using this extended target.
While the two charmonium states cannot be separated, it is estimated that
the $ \psi ^{\prime } $ contributes only about 1\% to the total
event count, based on the relative production cross sections and branching
ratios into muons. Therefore, the $ \psi ^{\prime } $ is not considered in 
the following discussion.

The mass distributions were plotted in bins in the Feynman-$ x $
variable $ x_{F} $, transverse momentum $ p_{T} $, and the dimuon
polar angle $ \vartheta  $ in the dimuon rest frame. We use the Collins-Soper 
frame~\cite{Collins:1977iv}.
This is identical to the Gottfried-Jackson frame~\cite{Gottfried:1964nx}, 
used in several earlier experiments, for $ p_{T}=0 $,
and to a very good approximation equivalent even at the highest
$ p_{T} $ values in this experiment.

The invariant mass distribution in each $(x_F,p_T)$-bin were fitted to a 
Gaussian peak plus an exponential
or polynomial background. All of the parameters, including the mass, were 
determined by the
fit. The number of events under the peak gave the combined $ J/\psi  $
and $ \psi ^{\prime } $ triple-differential, unnormalized cross
section in $ x_{F} $, $ p_{T} $, and $ \vartheta  $. Distributions as a 
function of $ \vartheta  $ were
then formed in each $ x_{F} $ and $ p_{T} $ bin.

The spectrometer and trigger acceptance was calculated with the help
of a Monte Carlo simulation of the $ J/\psi  $ production process,
which included all the measured magnetic fields and detector efficiencies
and geometry. The known properties of $ J/\psi  $ production from
previous experiments were used. However, because of the binning in
$ x_{F} $ and $ p_{T} $, exact knowledge of the form of the
cross section as a function of these two variables was not crucial.
Events were generated with a flat $ \vartheta  $ distribution,
corresponding to unpolarized production. Simulated data were passed
through the same analysis chain as the real data. The Monte Carlo
reproduced quite accurately the main features of the data, including
the mass and vertex resolutions.

The azimuthal-angle ($\phi$) dependence of the production cross section was
assumed to be flat in the Monte Carlo, since the two interacting hadrons
were unpolarized. The corresponding measured acceptance-corrected
distribution was essentially flat, consistent with a 2\% uncertainty
in the direction of the incoming beam. The decay distribution was
also assumed to be flat in $ \phi  $, consistent with previous
results~\cite{Biino:1987qu}. No attempt was made to extract the 
$ \phi  $ dependence of the decay in this analysis.
\begin{figure}[htb!]
{\centering \resizebox*{1\columnwidth}{!}{\includegraphics{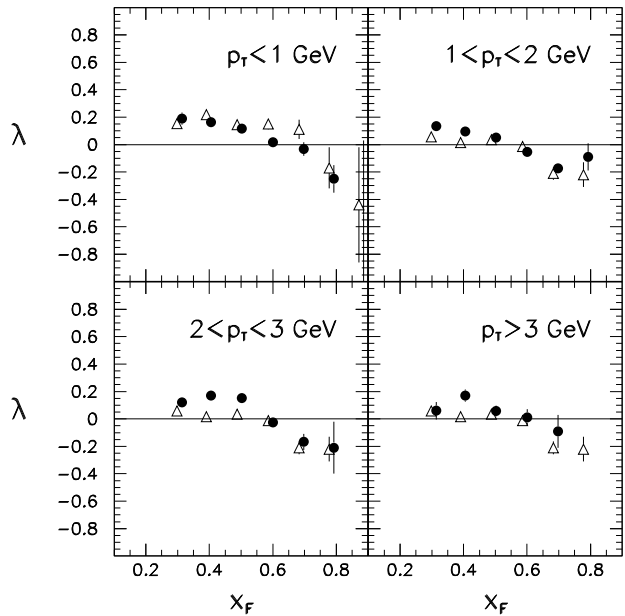}} \par}
\caption{$J/\Psi$ polarization parameter $\lambda$ 
versus \protect$ x_{F}\protect $ in \protect$ p_{T}\protect $
bins. Solid dots are the results obtained with the 2800 A magnet
setting, open triangles with the 2040 A setting. Only statistical
errors are shown.\label{lambda.xf.pt}}
\end{figure}

The $ \vartheta  $ distributions of the data in all bins were divided
by the corresponding ones of the simulation, resulting in acceptance-corrected
$ \vartheta  $ distributions. These were then used to calculate
the polarization in $ x_{F} $ and $ p_{T} $ bins, according
to the formula $ d\sigma /d\cos \vartheta =A(1+\lambda \cos ^{2}\vartheta ) $,
where the normalization constants $ A $ were left free. Results
were obtained separately for two experimental runs with the current
in the SM12 magnet set to 2040 and 2800 A respectively, resulting
in substantially different acceptances.

Figure \ref{lambda.xf.pt} shows the polarization parameter $ \lambda  $
as a function of $ x_{F} $ in four $ p_{T} $ bins; it can be
seen that results from the two magnet settings are in reasonable agreement, 
giving some confidence that the acceptance is understood and providing 
an estimate for the magnitude of the relevant systematic uncertainties.
The $ x_{F} $ dependence of $ \lambda $ appears to be independent of $ p_{T} $.
\begin{figure}[htb!]
{\centering \resizebox*{1\columnwidth}{!}{\includegraphics{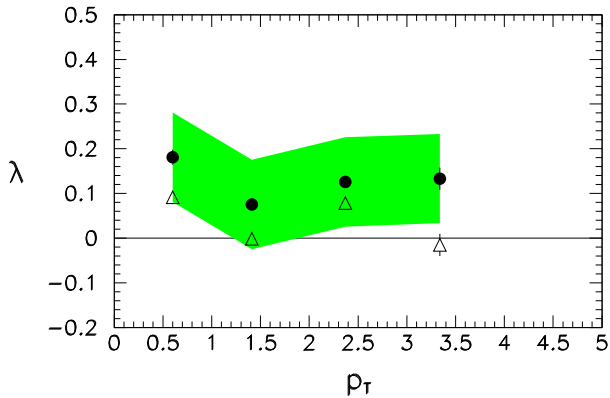}} \par}
\caption{$J/\Psi$ polarization parameter $\lambda$ 
versus \protect$ p_{T}\protect $ for two \protect$ x_{F}\protect $
ranges: \protect$ x_{F}<0.45\protect $ (solid circles) and \protect$ x_{F}>0.45\protect $
(open triangles). Statistical errors are smaller than the data points. 
Systematic errors for the small $ x_{F} $ data are shown as a dark band; 
those for large $ x_{F} $ (not shown) are slightly smaller.
\label{lambda.pt}}
\end{figure}

Systematic errors from various sources were considered.  Inexact knowledge
of the $ p_{T} $ dependence of the production cross section, coupled with
a strong $ p_{T} $ dependence of the acceptance versus decay angle $
\vartheta $, led to an uncertainty of $\pm0.06$ in $\lambda$, independent
of $x_F$.  Additional contributions included mass-peak fitting errors
(0.04--0.08, depending on the $x_F$ bin) and uncertainties in the exact
position (0.02) and angle (0.02--0.04) of the incoming beam and in the
fields in the analyzing magnets (0.01). Uncertainties from
various sources are largely uncorrelated; they were added in quadrature
for the overall systematic error.

Figure \ref{lambda.pt} presents the polarization parameter $ \lambda  $
as a function of $ p_{T} $ for two $ x_{F} $ ranges, where the
two data sets have been statistically combined. The two intervals
in $ x_{F} $ approximately correspond to regions were the gluon-gluon
and quark-antiquark processes are dominant. No significant $ p_{T} $
dependence is seen in either region after the systematic errors are
taken into account. At large $ p_{T} $, charmonium
production is understood~\cite{Braaten:1993rw} 
to be dominated by gluon bremsstrahlung with
subsequent fragmentation into a $ c\overline{c} $ pair. In this
case, the charmonium state is expected~\cite{Cho:1995ih,Braaten:1999qk} 
to retain to a large degree
the transverse polarization of the high-$ p_{T} $, on-shell gluon.
However, the $ p_{T} $ range of the experiment, extending to 4 
GeV, may not be sufficient to see clearly such an 
effect. It must be noted that this effect is also not observed 
at the high $ p_{T} $ values available at collider 
energies~\cite{Affolder:2000nn}, where, if anything, polarization appears 
to be longitudinal.
In the following, we assume there is no significant $ p_{T} $ dependence 
and the results are presented integrated over all $ p_{T} $. 

\begin{figure}[htb!]
{\centering \resizebox*{1\columnwidth}{!}{\includegraphics{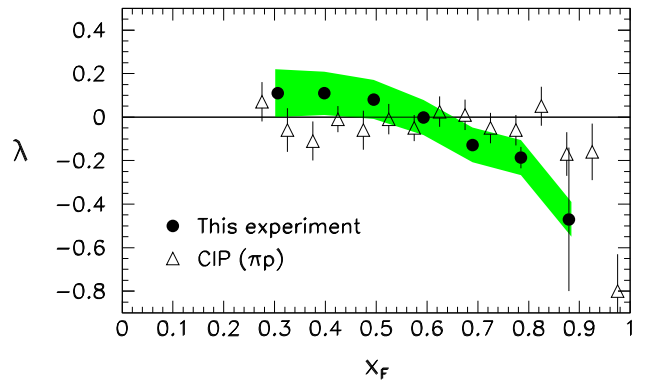}} \par}
\caption{\protect$ J/\psi \protect $ polarization parameter $\lambda$ versus \protect$ x_{F}\protect $
from this experiment (solid circles). Statistical errors are shown
as error bars, systematic as a dark band. Data values are tabulated 
in~\cite{Chang:1999hd}.
\label{lambda.xf} The CIP~\cite{Biino:1987qu} 
results are also shown (open triangles).}
\end{figure}

Figure \ref{lambda.xf} shows the polarization $ \lambda  $ as
a function of $ x_{F} $ for the combined data set. The one-sigma
systematic uncertainty is shown as a dark band. Also shown are the results
previously obtained by the
Chicago-Iowa-Princeton collaboration~\cite{Biino:1987qu}, which used a 252-GeV
pion beam and a tungsten target. Our results suggest 
that the $ J/\psi  $ is produced slightly transversely polarized
at small-to-intermediate $ x_{F} $. Within the systematic uncertainties
the results are consistent with the CIP experiment~\cite{Biino:1987qu}, 
which saw no polarization in this range, albeit with much larger statistical
uncertainties. It should be noted that in this range the $ q\overline{q} $
annihilation process is expected to play a more important role
with a pion beam than with a proton beam, where $ gg $ fusion 
dominates.
For $ x_{F}>0.6 $, the polarization turns to longitudinal, outside
the margins of the systematic error. This is similar to a pattern
seen in the CIP experiment, although the behavior appears smoother
as a function of $ x_{F} $ and begins at smaller $ x_{F} $ than the sudden 
turn-over near $ x_{F} \approx 0.85 $ in the latter.
However, the results are not incompatible. It is also interesting
to note that the clear change in polarization in this experiment roughly
coincides with the transition from gluon-fusion dominance to quark-annihilation
dominance.

Integrated over the entire $ x_{F} $ range, the measured polarization
is $ \lambda =0.069\pm 0.004 \pm 0.08 $ (statistical and systematic errors). This small
value, consistent with no polarization if systematic errors are taken
into account, is in agreement with previous proton-beam 
experiments~\cite{Alexopoulos:1997yd,Gribushin:1999ha},
which had insufficient statistics to study the $ x_{F} $ dependence.

As a cross-check of the analysis, the polarization of the Drell-Yan
continuum was also studied using the same technique. This can be done only
for dimuon invariant masses greater than 4 GeV, since the
$ J/\psi  $ peak dominates the spectrum at lower masses. An additional
complication arises from random coincidences of uncorrelated pairs
of muons, which in the case of the $ J/\psi  $ were removed by
the fitting procedure. These were estimated and subtracted from the
Drell-Yan sample by studying the distributions of same-charge muon
pairs. A fit to the formula $ 1+\lambda \cos ^{2}\vartheta  $ for dimuon 
masses from 4 to 7 GeV gave $ \lambda =0.98\pm 0.04$, in good
agreement with the 100\% transverse polarization expected 
and previously observed~\cite{Brown:2001bz} in the Drell-Yan
process. While the
kinematic range is not the same as for the $ J/\psi  $ measurement,
this agreement with expectations increases confidence in
the soundness of the $ J/\psi  $ analysis. The level of statistics were
not adequate to investigate any $ x_{F} $  dependence of the Drell-Yan 
polarization.

The results are not in agreement with published predictions for $ J/\psi  $
polarization based on NRQCD, which are in the range 
$ 0.31<\lambda <0.63 $~\cite{Beneke:1996tk}, or with
similar predictions~\cite{Vanttinen:1995sd} based on an early model
that considers only color-singlet intermediate states. 
The small positive values at $ x_{F} \lesssim 0.5 $ are very similar to
the results obtained for $ \Upsilon(1S) $ by this experiment in a similar
$ p_{T} $ range, and in sharp contrast to the essentially 100\% 
transverse polarization of the (unresolved) $ \Upsilon(2S) $ and
$ \Upsilon(3S) $ states~\cite{Brown:2001bz}. 

To understand the behavior of the polarization, all 
sources of $ J/\psi  $ production must be considered. While $ b $-quark
production is not a major source at these energies, almost half of
the produced $ J/\psi  $'s are the decay products of higher-mass
charmonium resonances, mainly the $ ^{3}P_{J} $ states $ \chi _{cJ} $.
Feed-down from $ \chi _{c2} $, produced exclusively in the $ J_{z}=2 $
state from gluon-gluon fusion, results in 100\% transverse $ J/\psi  $'s,
increasing the observed values of $ \lambda  $. The turn-over at
high $ x_{F} $ may reflect the transition, at $ x_{F} \simeq 0.6 $, 
to the quark-annihilation graph~\cite{Kowitt:1994ns}, 
which produces a mix of $ \chi _{c2} $ spin states. 
As for the overall level, it cannot be explained without a substantial contribution
from $ \chi _{c1} $, which in general produces longitudinal $ J/\psi  $'s.
Beneke and Rothstein~\cite{Beneke:1996tk} originally estimated the
$ \chi _{c1} $ contribution to be about 10 times smaller than that from 
$ \chi _{c2} $, but it was later found experimentally~\cite{Alexopoulos:1999wp} 
to be of comparable size. NRQCD calculations
which retain higher orders in $ v $ can also
accommodate substantial $ \chi _{c1} $ contributions~\cite{Beneke:1997jh}. 
It would
probably require a contribution near the experimental and theoretical
upper limits in order to obtain $ \lambda  $ values as small as
experimentally measured, within the context of NRQCD.

In addition, it must be noted that no existing calculation takes into
account nuclear effects, which strongly affect the production cross
section~\cite{Leitch:2000bz}. The formation length of the $ J/\psi $ 
at these energies is
generally longer than the nuclear size \cite{Farrar:1990ei}
and it is conceivable that color-singlet and -octet components 
of the wavefunction are absorbed differently while propagating through
the nuclear medium, resulting in a different mix of Fock states compared
to the free-nucleon production, thus altering the polarization.
Furthermore, the polarization of the $ \chi_c $ states themselves is 
predicted to have a nuclear dependence~\cite{Gerland:1998bz} which 
would feed down to the $ J/\psi $ polarization. Finally,
at the highest values of $ x_{F} $, higher-twist effects may become
important~\cite{Vanttinen:1995sd}.

\begin{acknowledgments}
This work was supported in part by the U.S.~Department of Energy.
\end{acknowledgments}

\bibliography{Jpsi}

\end{document}